\documentclass[]{spie}

\usepackage{graphicx}
\usepackage{algorithm}
\usepackage{algorithmic}
\usepackage{amsmath}
\usepackage{amsfonts}
\usepackage{subfigure}
\usepackage{booktabs}
\usepackage{multirow}
\graphicspath{{./figs/}}

\title{Triple Patterning Lithography (TPL) Layout Decomposition using End-Cutting} 

\author{Bei Yu, Jhih-Rong Gao, and David Z. Pan
\skiplinehalf
ECE Dept. University of Texas at Austin, Austin, TX USA 78712 \\
Email: \{bei, jrgao, dpan\}@cerc.utexas.edu
}

\begin{document}
\maketitle

\vspace{0.5in}
\begin{abstract}

Triple patterning lithography (TPL) is one of the most promising techniques in the 14nm logic node and beyond.
However, traditional LELELE type TPL technology suffers from native conflict and overlapping problems.
Recently LELEEC process was proposed to overcome the limitations, where the third mask is used to generate the end-cuts.
In this paper we propose the first study for LELEEC layout decomposition.
Conflict graphs and end-cut graphs are constructed to extract all the geometrical relationships of input layout and end-cut candidates.
Based on these graphs, integer linear programming (ILP) is formulated to minimize the conflict number and the stitch number.

\end{abstract}

\section{Introduction}

As the semiconductor process further scales down, the industry encounters many lithography-related issues.
In the 14nm logic node and beyond, triple patterning lithography (TPL) \cite{TPL_SPIE2012_Lucas} is one of the most promising techniques \cite{LITH_ICCAD2012_Yu},
because of the delay of other candidates, such as extreme ultraviolet lithography (EUVL) and electron beam lithography (EBL).
EUVL is challenged by tremendous technical barriers \cite{EUV_SPIE2010_Arisawa},
while EBL has a serious limitation due to low throughput \cite{EBL_TCAD2012_Yuan, EBL_ASPDAC2013_Yu}.

\begin{figure}[tb]
  \centering
  \hspace{.16in}
  \subfigure[]{\includegraphics[width=0.18\textwidth]{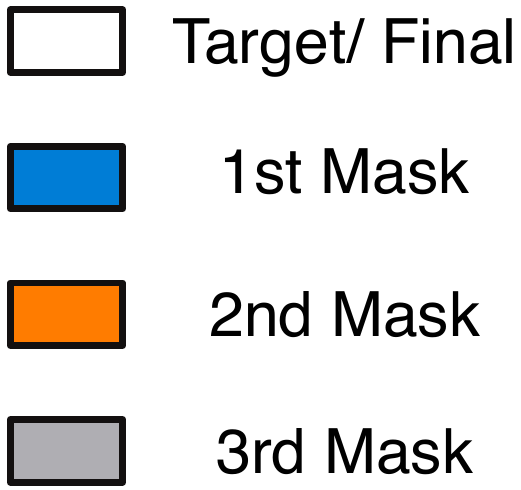}}
  \hspace{.16in}
  \subfigure[]{\includegraphics[width=0.24\textwidth]{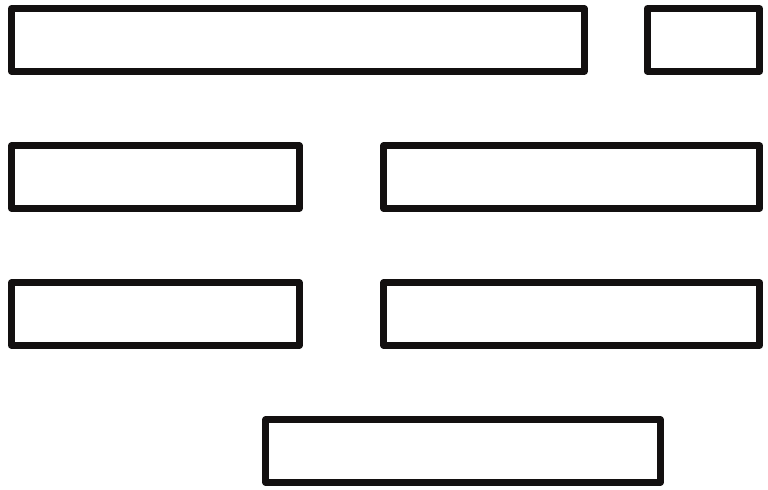}}
  \hspace{.1in}
  \subfigure[]{\includegraphics[width=0.24\textwidth]{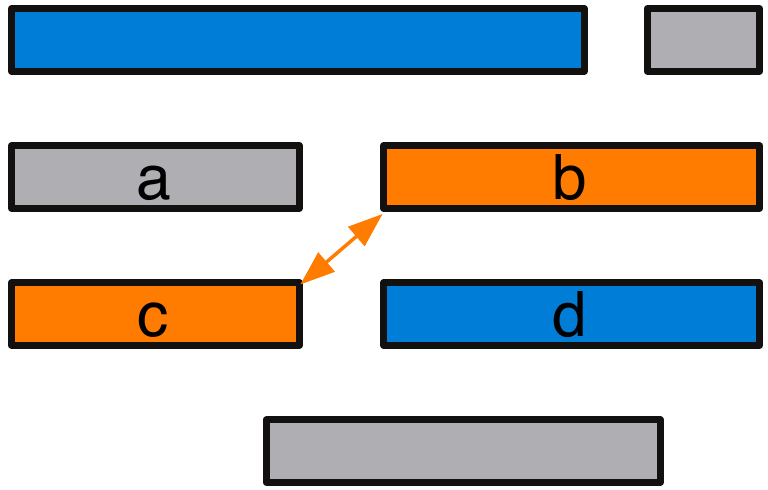}}
  \caption{Process of LELELE~(a) Some labels.~(b) Target features.~(c) layout decomposition with one conflict introduced.}
  \label{fig:LELELE}
\end{figure}

One traditional process of TPL is with the same principle of litho-etch-litho-etch (LELE) type double patterning lithography (DPL),
where the original layout is decomposed into three masks and manufactured through three exposure/etching steps.
This technology is called LELE-litho-etch (LELELE).
Many research has been carried out to solve the corresponding design problems, including layout decomposition
\cite{TPL_SPIE08_Cork,TPL_ICCAD2011_Yu,TPL_SPIE2011_Ghaida,TPL_DAC2012_Fang,TPL_ICCAD2012_Tian} and routing \cite{DFM_DAC2012_Ma,DFM_ICCAD2012_Lin}.
However, even with stitch insertion, there are some native conflicts in LELELE.
Fig. \ref{fig:LELELE} shows a 4-clique conflict among features $a$, $b$, $c$, and $d$.
Since this 4-clique structure is common in advanced standard cell design, LELELE type TPL still suffers from the native conflict problem \cite{TPL_ICCAD2011_Yu}.
Besides, compared with LELE, there are more serious overlapping problem in LELELE \cite{TPL_SPIE08_Ausschnitt}.

\begin{figure}[tb]
  \centering
  \subfigure[]{\includegraphics[width=0.18\textwidth]{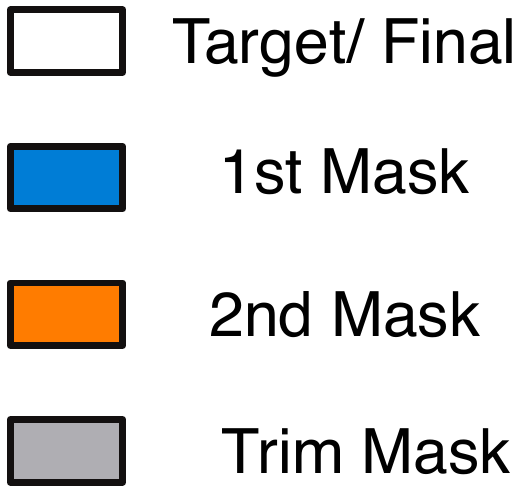}}
  \subfigure[]{\includegraphics[width=0.24\textwidth]{1Target}}
  \hspace{.1in}
  \subfigure[]{\includegraphics[width=0.24\textwidth]{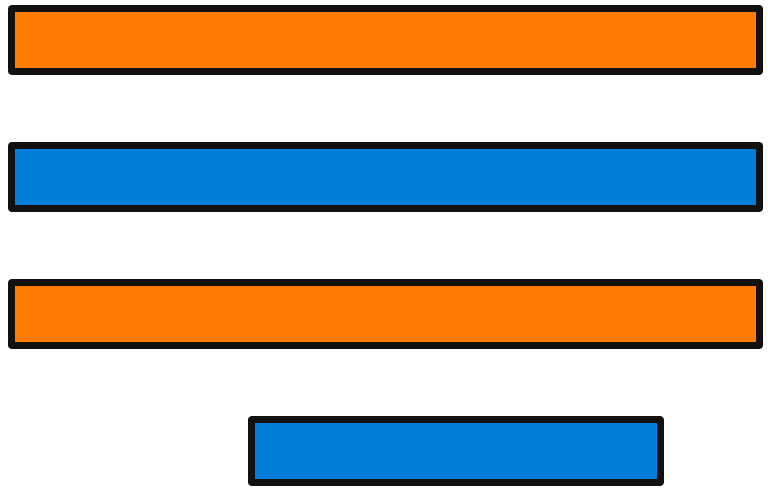}}
  \hspace{.1in}
  \subfigure[]{\includegraphics[width=0.24\textwidth]{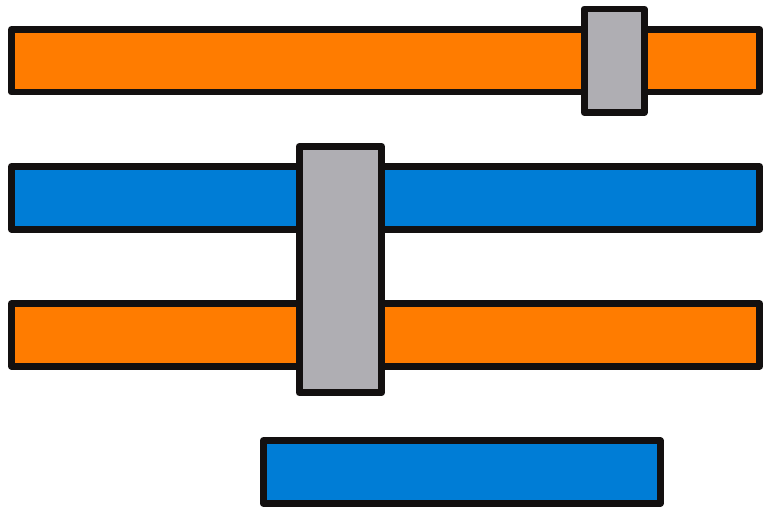}}
  \caption{Process of LELEEC~(a) Some labels.~(b) Target features.~(c) First and second mask patterns.~(d) Trim mask, and final decomposition with no conflict.}
  \label{fig:LELEEC}
\end{figure}

To overcome all these limitations derived from LELELE, recently Lin \cite{LITH_ISPD2012_Lin} proposes a new TPL manufacturing process, called LELE-end-cutting (LELEEC).
As a TPL, this new manufacturing process contains three mask steps, namely first mask, second mask, and \textit{trim} mask.
Fig. \ref{fig:LELEEC} illustrates an example of the LELEEC process.
To generate target features in Fig. \ref{fig:LELEEC}(b), the first and second masks are used for pitch splitting,
which is similar to LELE process in DPL.
These two masks are shown in Fig. \ref{fig:LELEEC}(c).
Finally, a trim mask is used to trim out the desired region as shown in Fig. \ref{fig:LELEEC}(d).
In other words, the trim mask is used to generate some end-cuts to further split feature patterns.
As a result, the features that are not LELELE-friendly can be printed without introducing any conflict through the LELEEC process.
Besides, the end-cuts introduce better printability \cite{TPL_ISPD2012_Lin}.

Layout decomposition, which is the most crucial step for TPL, has been extremely studied under LELELE process 
\cite{TPL_SPIE08_Cork,TPL_ICCAD2011_Yu,TPL_SPIE2011_Ghaida,TPL_DAC2012_Fang}.
However, some new design challenges are introduced in LELEEC process that previous layout decomposition methodology cannot be directly borrowed.
Since the end-cuts are printed through one mask, some end-cuts may not be compatible to each other.
How to design the trim mask is pretty crucial, and therefore the layout decomposition for LELEEC is still an open problem.

In this paper, we propose the first study for LELEEC layout decomposition.
Given a layout which is specified by features in polygonal shapes, we extract the geometrical relationships and construct the conflict graphs.
Furthermore, the compatibility of all end-cuts candidates are also modeled in the conflict graphs.
Based on the conflict graphs, integer linear programming (ILP) is formulated to assign each vertex into one layer.
Our goal in the layout decomposition is to minimize the conflict number, and at the same time minimize the overlapping errors.

The rest of the paper is organized as follows. In Section \ref{sec:problem}, we provide some preliminary, and discuss the problem formulation.
In Section \ref{sec:endcut} we explain the details to generate the end-cut candidates.
In Section \ref{sec:algo} we provide the integer linear programming formulation for this problem, and followed by several speedup techniques in Section \ref{sec:speedup}.
Section \ref{sec:result} presents the experimental results, and we conclude this paper in Section \ref{sec:conclu}.

\section{Preliminary and Problem Formulation}
\label{sec:problem}

\subsection{Layout Graph}

\begin{figure}[htb]
  \centering
   \subfigure[]{\includegraphics[width=0.3\textwidth]{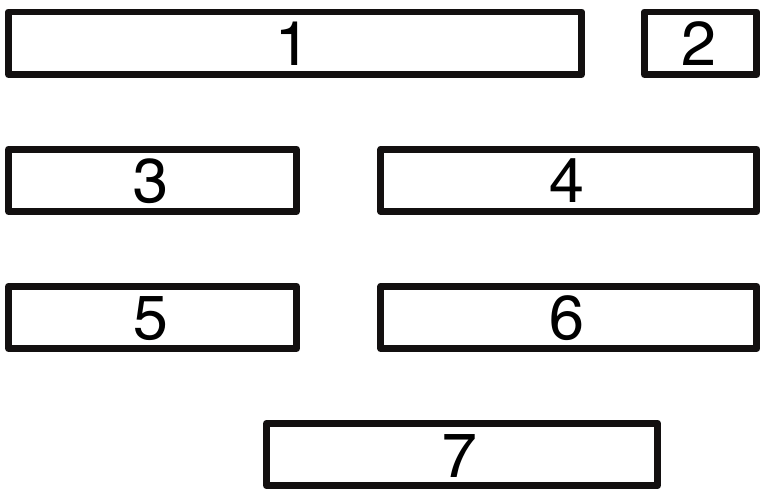}}
   \hspace{.16in}
   \subfigure[]{\includegraphics[width=0.26\textwidth]{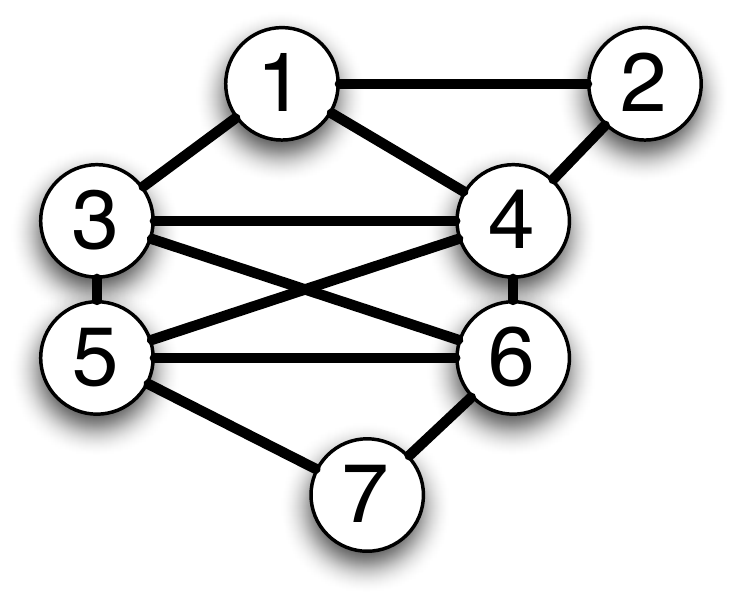}}\\
   \hspace{.2in}
   \subfigure[]{\includegraphics[width=0.32\textwidth]{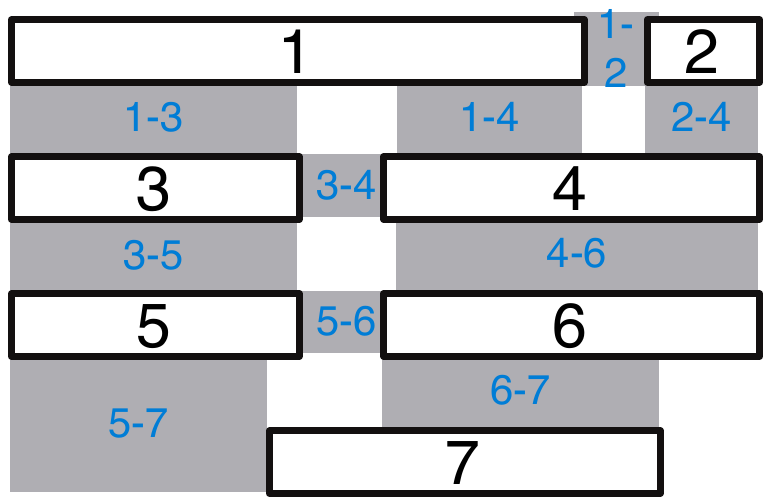}}
   \hspace{.26in}
   \subfigure[]{\includegraphics[width=0.34\textwidth]{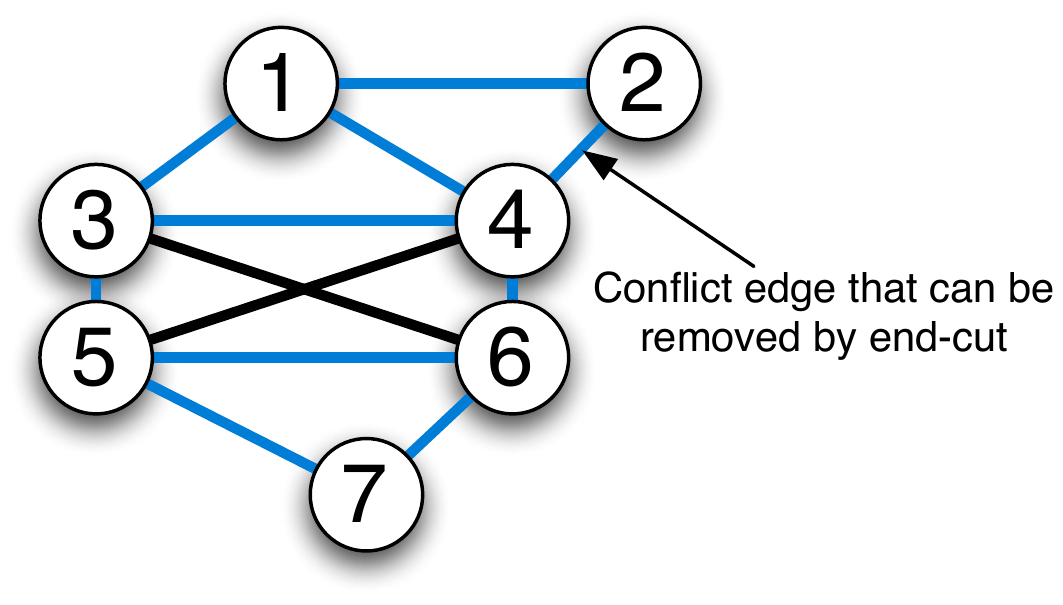}}
  \caption{Layout graph construction~(a) Input layout.~(b) Layout graph with conflict edges.~(c) End-cut candidate generation, where each grey rectangle represents one end-cut candidate.~(d) Update layout graph where each blue edge means that the conflict can be removed by introducing one end-cut.}
  \label{fig:LG}
\end{figure}

Given a layout which is specified by features in polygonal shapes, layout graphs \cite{TPL_ICCAD2011_Yu} are constructed.
As shown in Fig. \ref{fig:LG}, the layout graph is an undirected graph with a set of vertices $V$ and a set of conflict edges $CE$.
Each vertex in $V$ represents one input feature.
There is an edge in $CE$ if and only if the two features are within minimum coloring distance $dis_m$ of each other.
In other words, each edge in $CE$ is a conflict candidate.
Fig. \ref{fig:LG}(a) shows one input layout, and the corresponding layout graph is in Fig. \ref{fig:LG}(b).
For each edge (conflict candidate), we check whether there is an end-cut candidate.
End-cut candidates (grey rectangles in Fig. \ref{fig:LG}(c)) are introduced to input layout.
For each end-cut candidate $i-j$, if it is applied, then features $i$ and $j$ will be merged into one feature.
By this way the corresponding conflict edge can be removed.
We label all edges in layout graph, to indicate those conflict edges that can be removed by end-cut insertion (see blue solid edges in Fig. \ref{fig:LG}(d)).
If stitch is considered in layout decomposition, some vertices in layout graph can be split into several segments.
The segments in one layout graph vertex are connected through \textit{stitch edges}.
A set $SE$ contains all these stitch edges \footnote{Please refer to \cite{DPL_ICCAD08_Kahng} for the details of stitch candidate generation.}.

\subsection{End-Cut Graph}

\begin{figure}[htb]
  \centering
  \subfigure[]{\includegraphics[width=0.32\textwidth]{LG3}}
  \hspace{.2in}
  \subfigure[]{\includegraphics[width=0.5\textwidth]{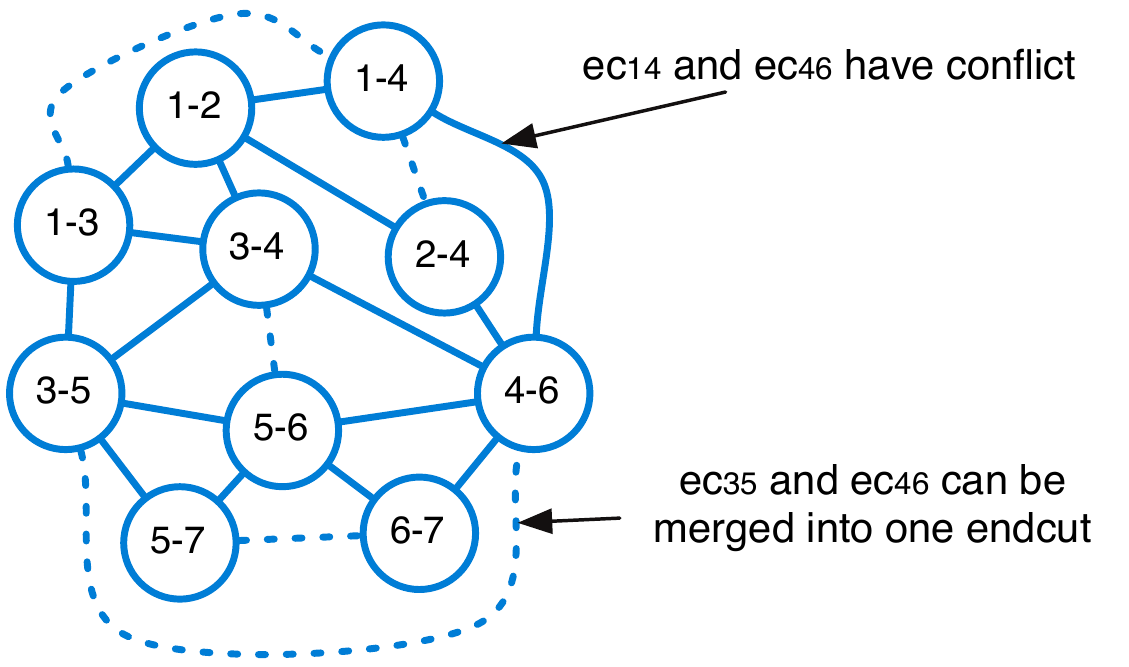}}
  \caption{End-cut graph construction.~(a) Input layout with end-cut candidates.~(b) End-cut graph.}
  \label{fig:EC_graph}
\end{figure}

Since all the end-cuts are manufactured through one single exposure process, they should be distributed far away from each other.
That is, two end-cuts have conflicts if they are within minimum end-cut distance $dis_c$ of each other.
Note that these conflict relationships among end-cuts are not available in layout graph, therefore we construct \textit{end-cut graph} to store the relationships.
Fig. \ref{fig:EC_graph}(a) shows an example of an input layout with all end-cut candidates;
the corresponding end-cut graph is shown in Fig. \ref{fig:EC_graph}(b).
Each vertex in the graph represents one end-cut.
There is an solid edge if and only if the two end-cuts conflict to each other.
There is an dash edge if and only if they are close to each other, and they can be merged into one larger end-cut. 

\subsection{Problem Formulation}

Given a layout which is specified by features in polygonal shapes, the layout graph and the end-cut graph are constructed.
The layout decomposition assigns all vertices in layout graph into one of two colors, and select a set of end-cuts in end-cut graph.
The objectives it to minimize the number of conflict and/or stitch.

\section{End-Cut Candidate Generation}
\label{sec:endcut}

\begin{figure}[htb]
  \centering
 \includegraphics[width=0.95\textwidth]{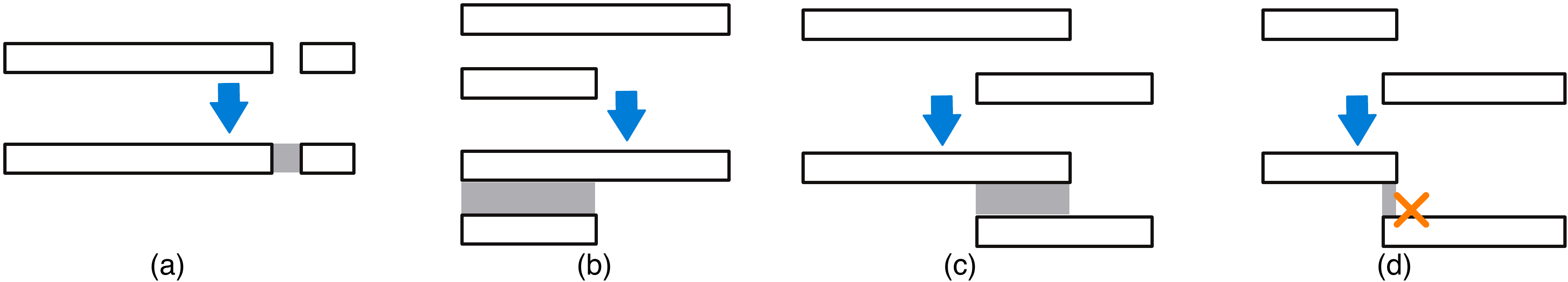}
  \caption{End-cut candidate generation for edge-edge type.}
  \label{fig:candidate_e2e}
\end{figure}

Above we have discussed how to generate the layout graph and end-cut graph.
In this section we will explain the details to generate all end-cut candidates.
For each conflict edge in layout graph, there are two adjacent features in corresponding input layout.
To check whether one end-cut can be inserted, we classify the adjacent relationships of features into two types:
\textit{edge-edge} and \textit{corner-corner}.
The end-cut between two edge-edge features can be calculated through calculating the projection, as illustrated in Fig. \ref{fig:candidate_e2e}.
Note that in some cases, if the width of one end-cut is smaller than wire width threshold $w_{th}$, this end-cut candidate is forbidden (see Fig. \ref{fig:candidate_e2e}(d)). 

\begin{figure}[htb]
  \centering
  \includegraphics[width=0.5\textwidth]{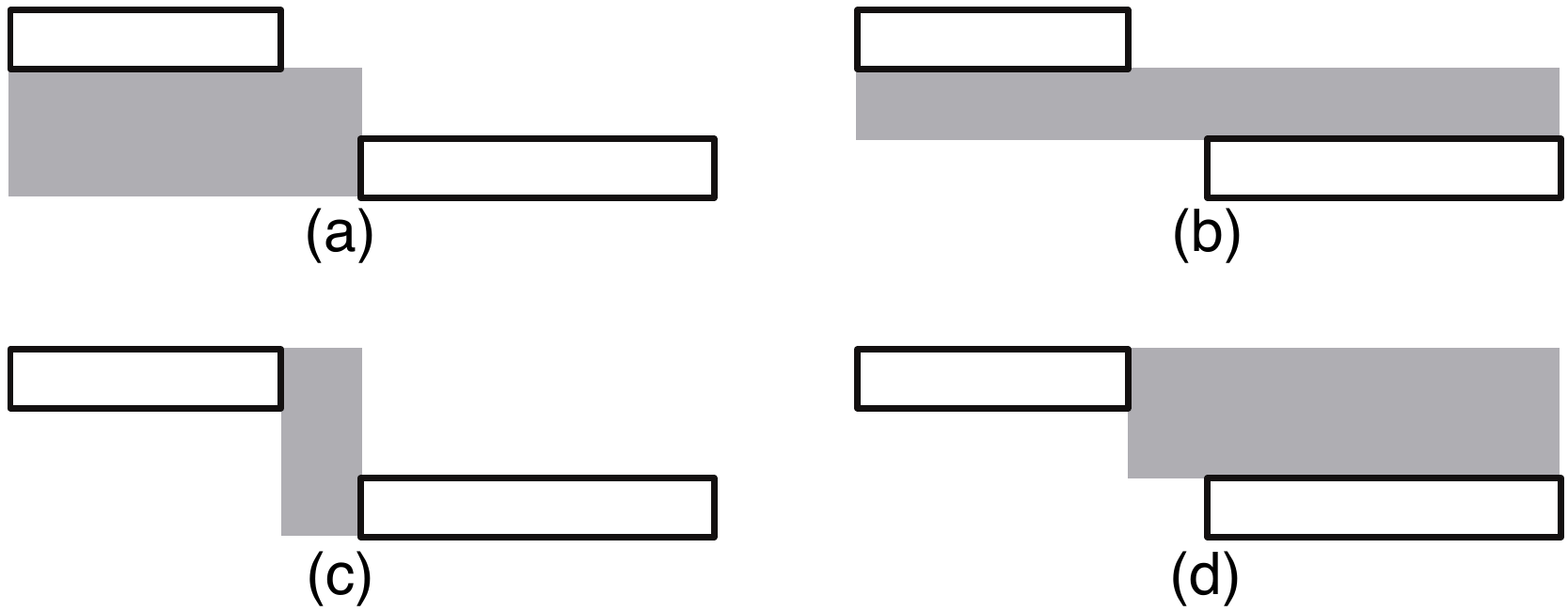}
  \caption{End-cut candidate generation for corner-corner type.}
  \label{fig:candidate_c2c}
\end{figure}

If one end-cut candidate cannot be constructed through edge-edge type, we search whether there is one legal corner-corner type end-cut.
As shown in Fig. \ref{fig:candidate_c2c}, there are four possible end-cut shapes.
We pick up the shape with minimal area, meanwhile the shape width should be larger than width threshold $w_{th}$.
For each end-cut candidate, if it is overlapping with any layout feature, it would be forbidden.

\section{ILP Algorithms}
\label{sec:algo}

After the construction of layout graph and end-cut graph, LELEEC layout decomposition problem can be transferred into an optimization problem on graphs.
Note that conflict graph can not be guaranteed to be planar, so some face based methodology \cite{DPL_ISPD2010_Xu} cannot be applied here.
Therefore, we formulate integer linear programming (ILP) to solve the optimization problem.
For convenience, some notations in the ILP formulation are listed in Table \ref{table:ilp}.

\begin{table}[hbt]
\renewcommand{\arraystretch}{1.1}
\centering 
\caption{Notations in ILP}
\label{table:ilp}
\begin{tabular}{|c|c|}
	\hline \hline
	$CE$		    & set of conflict edges\\
	\hline
	$EE$		    & set of end-cut conflict edges\\
  \hline
	$SE$		    & set of stitch edges.\\
	\hline
	$r_i$		    & the $i_{th}$ layout feature\\
	\hline
	$x_i$		    & variable denoting the coloring of $r_i$\\
	\hline
  $ec_{ij}$   & 0-1 variable, $ec_{ij}=1$ when the end-cut between $r_i$ and $r_j$\\
  \hline
	$c_{ij}$	  & 0-1 variable, $c_{ij}=1$ when a conflict between $r_i$ and $r_j$\\
	\hline
	$s_{ij}$	  & 0-1 variable, $s_{ij}=1$ when a stitch between $r_i$ and $r_j$\\
	\hline\hline
\end{tabular}
\end{table}

\subsection{ILP Formulation without Stitch Insertion}

When no stitch candidate is considered, the layout decomposition problem can be formulated as shown in Eq. (\ref{eq:ILP}).
The objective is to minimize the conflict number.

\begin{align}
  \textrm{min}      &  \sum_{e_{ij} \in CE} c_{ij}  &   \label{eq:ILP}\\
  \textrm{s.t.} \ \ &  x_i + x_j \le 1 + c_{ij} + ec_{ij}           & \forall e_{ij} \in CE \label{ILPa}\tag{$1a$} \\
                    &  (1-x_i) + (1-x_j) \le 1 + c_{ij} + ec_{ij}   & \forall e_{ij} \in CE \label{ILPb}\tag{$1b$} \\
                    & ec_{ij} + ec_{pq} \le 1                       & \forall e_{ijpq} \in EE \label{ILPc}\tag{$1c$} \\
                    & ec_{ij} + x_i - x_j \le 1                     & \forall e_{ij} \in CE \label{ILPd}\tag{$1d$}\\
                    & ec_{ij} + x_j - x_i \le 1                     & \forall e_{ij} \in CE \label{ILPe}\tag{$1e$}
\end{align}
where $x_i$ is a variable for the colors of feature $r_i$, $c_{ij}$ is a binary variable for conflict edge $e_{ij} \in CE$.
$ec_{ij}$ is a binary variable for end-cut candidate.
Constraints (\ref{ILPa}) and (\ref{ILPb}) are used to evaluate the conflict number and end-cut number.
If two adjacent features $r_i$ and $r_j$ are assigned same colors (masks), and the end-cut candidate $ec_{ij}$ is not applied ($ec_{ij} = 0$),
then there is one conflict reported ($c_{ij} = 1$).
If end-cuts $ec_{ij}$ and $ec_{pq}$ are conflict with each other, constraint (\ref{ILPc}) makes sure that at most one of them will be applied.
Constraints (\ref{ILPd}) and (\ref{ILPe}) are used to forbid useless end-cut.
That is, if features $x_{i}$ and $x_{j}$ are in different colors, $ec_{ij} = 0$.

\begin{figure}[htb]
  \centering
  \includegraphics[width=0.65\textwidth]{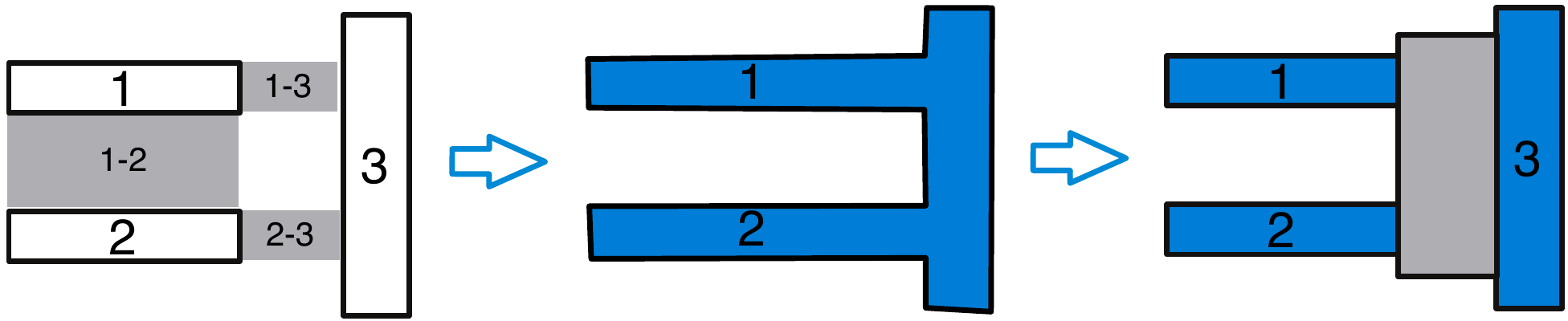}
  \caption{End-cut candidate generation for corner-corner type.}
  \label{fig:ilp_error}
\end{figure}

At first glance, the ILP formulation (\ref{eq:ILP}) works well.
However, it may report some unnecessary conflicts.
An example is shown in Fig. \ref{fig:ilp_error}, where three end-cut candidates $ec_{12}, ec_{13}, ec_{23}$ are considered among features $r_1, r_2$ and $r_3$.
Note that $ec_{13}$ and $ec_{23}$ could be merged into one large end-cut.
If these two end-cuts are applied, all features can be manufactured through one feature (see Fig. \ref{fig:ilp_error}(b)).
However, based on the constraints (\ref{ILPa}) and (\ref{ILPb}), since $x_1$ equals to $x_2$, and end-cut $ec_{12}$ is not applied, conflict $c_{12}$ will be reported.
We can see that our initial ILP formulation ignore this wire merging character.
To overcome this limitation, we modify the constraints (\ref{ILPa}) and (\ref{ILPb}) as follows:

\begin{align}
  x_i + x_j \le 1 + c_{ij} + ec_{ij} + ec_{ik} \cdot ec_{jk}          \label{eq:times}\\
  (1-x_i) + (1-x_j) \le 1 + c_{ij} + ec_{ij} + ec_{ik} \cdot ec_{jk}  \notag 
\end{align}

Since $ec_{ik} \cdot ec_{jk}$ is non-linear, we introduce boolean variables $\gamma_{ik, jk}$ to replace it, and enforce the following artificial constraints in our ILP formulation:
\begin{align}
  ec_{ik} + ec_{jk} \le \gamma_{ik, jk} + 1      \notag\\
  ec_{ik} \ge \gamma_{ik, jk}                   \label{eq:linear}\\
  ec_{jk} \ge \gamma_{ik, jk}                   \notag
\end{align}

After replacing constraints (\ref{eq:times}) with (\ref{eq:linear}), our formulation can be linearized.
Therefore, the modified ILP formulation is as follows:

\begin{align}
  \textrm{min}      &  \sum_{e_{ij} \in CE} c_{ij}  &   \label{eq:ILP2}\\
  \textrm{s.t.} \ \ &  x_i + x_j \le 1 + c_{ij} + ec_{ij} + \gamma_{ik,jk}             & \forall e_{ij} \in CE \tag{$4a$} \\
                    &  (1-x_i) + (1-x_j) \le 1 + c_{ij} + ec_{ij} + \gamma_{ik,jk}     & \forall e_{ij} \in CE \tag{$4b$} \\
                    &  (1c) - (1e), (3)                                                & \notag
\end{align}

\subsection{ILP formulation with Stitch Insertion}

If the stitch insertion is considered, the ILP formulation is as in Eq. (\ref{eq:ILP3}).
The objective is to simultaneously minimize both the conflict number and the stitch number.
The parameter $\alpha$ is a user-defined parameter for assigning relative importance between the conflict number and the stitch number.

\begin{align}
  \textrm{min} & \sum_{e_{ij}\in CE} c_{ij} + \alpha \times \sum_{e_{ij} \in SE} s_{ij} & \label{eq:ILP3} \\
  \textrm{s.t.} \ \ &  x_i - x_j \le s_{ij}                          & \forall e_{ij} \in SE \tag{$5a$} \\
                    &  x_j - x_i \le s_{ij}                          & \forall e_{ij} \in SE \tag{$5b$} \\
                    &  (1c) - (1e), (3), (4a) - (4b)                 & \notag
\end{align}

\section{Speedup Techniques}
\label{sec:speedup}

ILP is a well-known NP-hard problem that it may suffer from long runtime penalty to achieve the results.
In this section, we provide a set of speedup techniques.
Note that these techniques can keep optimality.
In other words, with these speedup techniques, ILP formulation can achieve the same results comparing with those not applying speedup.

\subsection{Independent Component Computation}

The first speedup technique is called \textit{independent component computation}.
By breaking down the whole layout graph into several independent components, we partition the initial layout graph into several small ones.
Then each component can be resolved through ILP formulation independently.
At last, the overall solution can be taken as the union of all the components without affecting the global optimality.
Note that this is a well-known technique which has been applied in many previous studies ( e.g., \cite{DPL_ICCAD08_Kahng,DPL_ISPD09_Yuan,DPL_ASPDAC2010_Yang}).

\subsection{Bridge Computation}

A bridge of a graph is an edge whose removal disconnects the graph into two components.
If the two components are independent, removing the bridge can divide the whole problem into two independent sub-problems.
We search all bridge edges in layout graph, then divide the whole layout graph through these bridges.
Note that all bridge can be found in $O(|V|+|E|)$, where $|V|$ is the vertex number, and $|E|$ is the edge number in the layout graph.

\subsection{End-Cut Pre-Selection}

Some end-cut candidates have no conflict end-cuts.
For the end-cut candidate $ec_{ij}$ that has no conflict end-cut, it would be pre-selected in final decomposition results.
That is, the features $r_i$ and $r_j$ are merged into one feature.
By this way, the problem size of ILP formulation can be further reduced.

\section{Experimental Results}
\label{sec:result}

We implement our algorithms in C++ and test on an Intel Xeon 3.0GHz Linux machine with 32G RAM.
ISCAS 85\&89 benchmarks from \cite{TPL_ICCAD2011_Yu} are used.
GUROBI \cite{Gurobi} is chosen as the ILP solver.
We set the minimum coloring distance $dis_m$ as $2W_{min}+3S_{min}$, where $W_{min}$ and $S_{min}$ denote the minimum wire width and the minimum spacing, respectively.
The width threshold $w_{th}$, which is used in end-cut candidate generation, is set as $dis_m$.

\subsection{With or Without Stitch}

\begin{table*}[htb]
\centering
\renewcommand{\arraystretch}{0.9}
\caption{Comparison of Runtime and Performance}
\label{tab:result1}
\begin{tabular}{|c|c|c |c|c|c|c||c|c|c|c||}
  \hline \hline
  \multirow{2}{*}{Circuit} &\multirow{2}{*}{Wire\#}  &\multirow{2}{*}{Sub-G\#} &\multicolumn{4}{c||}{ILP w/o. stitch} & \multicolumn{4}{c|}{ILP w. stitch}\\
  \cline{4-11}
            &          &       &conflict\# &stitch\#  & cost & CPU(s)                 &conflict\#    &stitch\#      &cost  & CPU(s) \\
  \hline
  C432      &1109      &30     &0          &0         &0    &1.66                 &0   &0     &0    &1.77        \\
  C499      &2216      &64     &0          &0         &0    &3.7                  &0   &0     &0    &4.15        \\
  C880      &2411      &102    &0          &0         &0    &5.3                  &0   &0     &0    &5.56        \\
  C1355     &3262      &104    &0          &0         &0    &5.5                  &0   &0     &0    &5.72        \\
  C1908     &5125      &155    &1          &0         &1    &8.4                  &0   &0     &0    &8.62        \\
  C2670     &7933      &299    &0          &0         &0    &16.2                 &0   &0     &0    &16.75       \\
  C3540     &10189     &417    &0          &0         &0    &22.5                 &0   &0     &0    &23.33       \\
  C5315     &14603     &601    &0          &0         &0    &32.8                 &0   &0     &0    &34.01       \\
  C6288     &14575     &475    &9          &0         &9    &27.5                 &8   &11    &9.1  &32.54       \\
  C7552     &21253     &788    &0          &0         &0    &44.3                 &0   &0     &0    &45.57       \\
  S1488     &4611      &194    &0          &0         &0    &10.3                 &0   &0     &0    &10.73       \\
  S38417    &67696     &2285   &3          &0         &3    &149.6                &0   &1     &0.1  &159.68      \\
  S35932    &157455    &4469   &47         &0         &47   &411.3                &5   &1     &5.1  &380.73      \\
  S38584    &168319    &5659   &14         &0         &14   &502.3                &2   &0     &2    &477.27      \\
  S15850    &159952    &5417   &16         &0         &16   &473.9                &5   &4     &5.4  &452.58      \\
  \hline
  avg.      &-         &-      & 6         &0         &6    &114.3                & 1.33& 1.13& 1.45&110.6      \\
  \hline
  ratio     &-         &-      & 4.5	   &0  &\textbf{1.0}  &\textbf{1.0}     & 1.0 &1.0 &\textbf{0.24} &\textbf{0.97}\\
  \hline \hline
\end{tabular}
\end{table*}

In the first experiment, we show the decomposition results of the ILP formulation.
Table \ref{tab:result1} compares two ILP formulations ``ILP w/o. stitch'' and ``ILP w. stitch''.
Here ``ILP w/o. stitch'' is the ILP formulation based on the graph without stitch edges, while ``ILP w. stitch'' considers the stitch insertion in the ILP.
Columns ``Wire\#'' and ``Sug-G\#'' reports the total feature number, and the divided sub-graph number, respectively.
For each method we report the conflict number, stitch number, and computational time in seconds(``cpu(s)'').
``Cost'' is the cost function, which is set as conflict\# $+ 0.1 \times $ stitch\#.
From Table \ref{tab:result1} we can see that both ILP formulation is effective that only a few conflicts are reported.
Compared with ``ILP w/o. stitch'', when stitch candidates are considered in the ILP formulation, the cost can be reduced by 76\%, while the runtime is similar.

\begin{figure}[htb]
  \centering
  \subfigure[]{\includegraphics[width=0.47\textwidth]{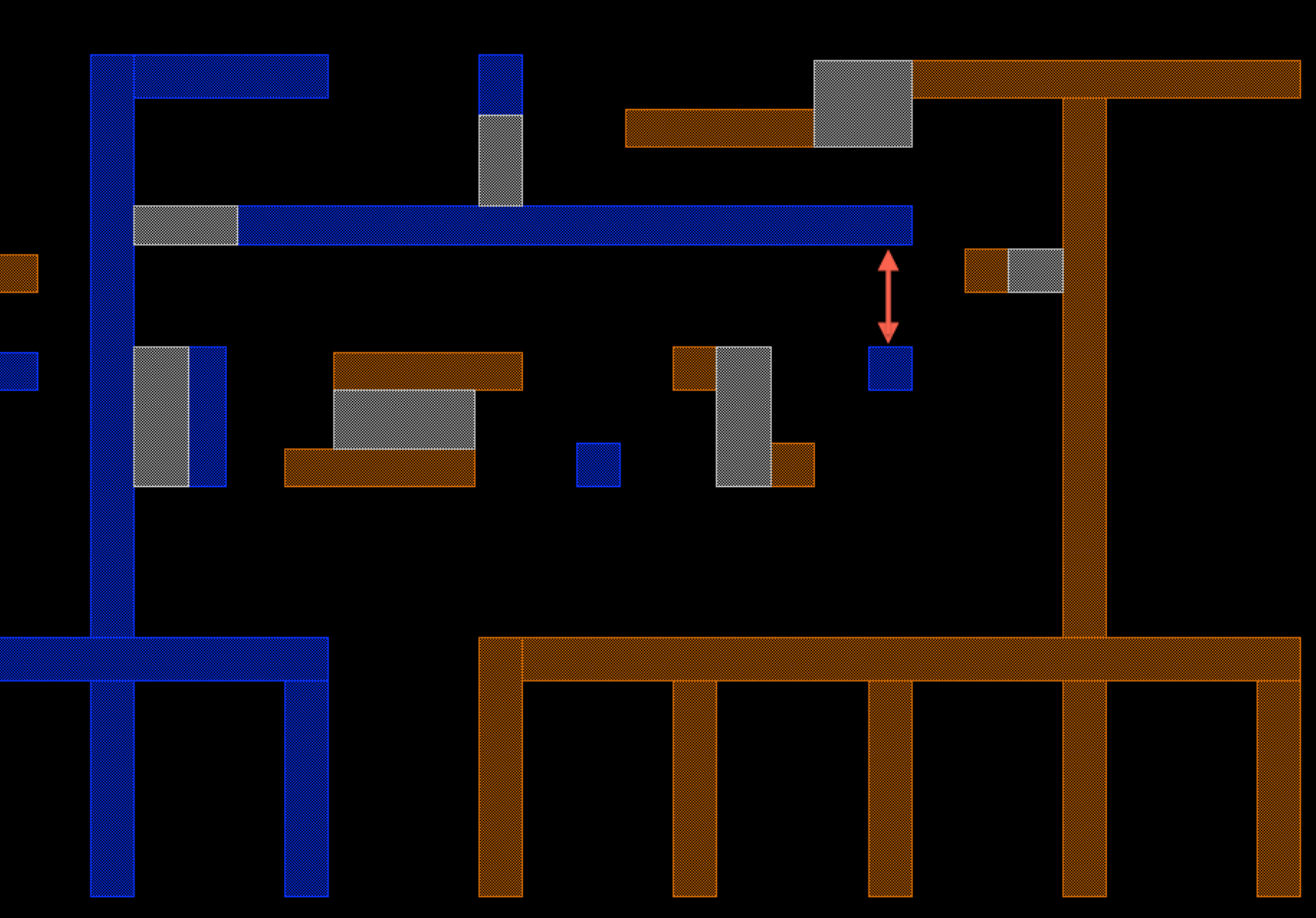}}
  \hspace{.2in}
  \subfigure[]{\includegraphics[width=0.45\textwidth]{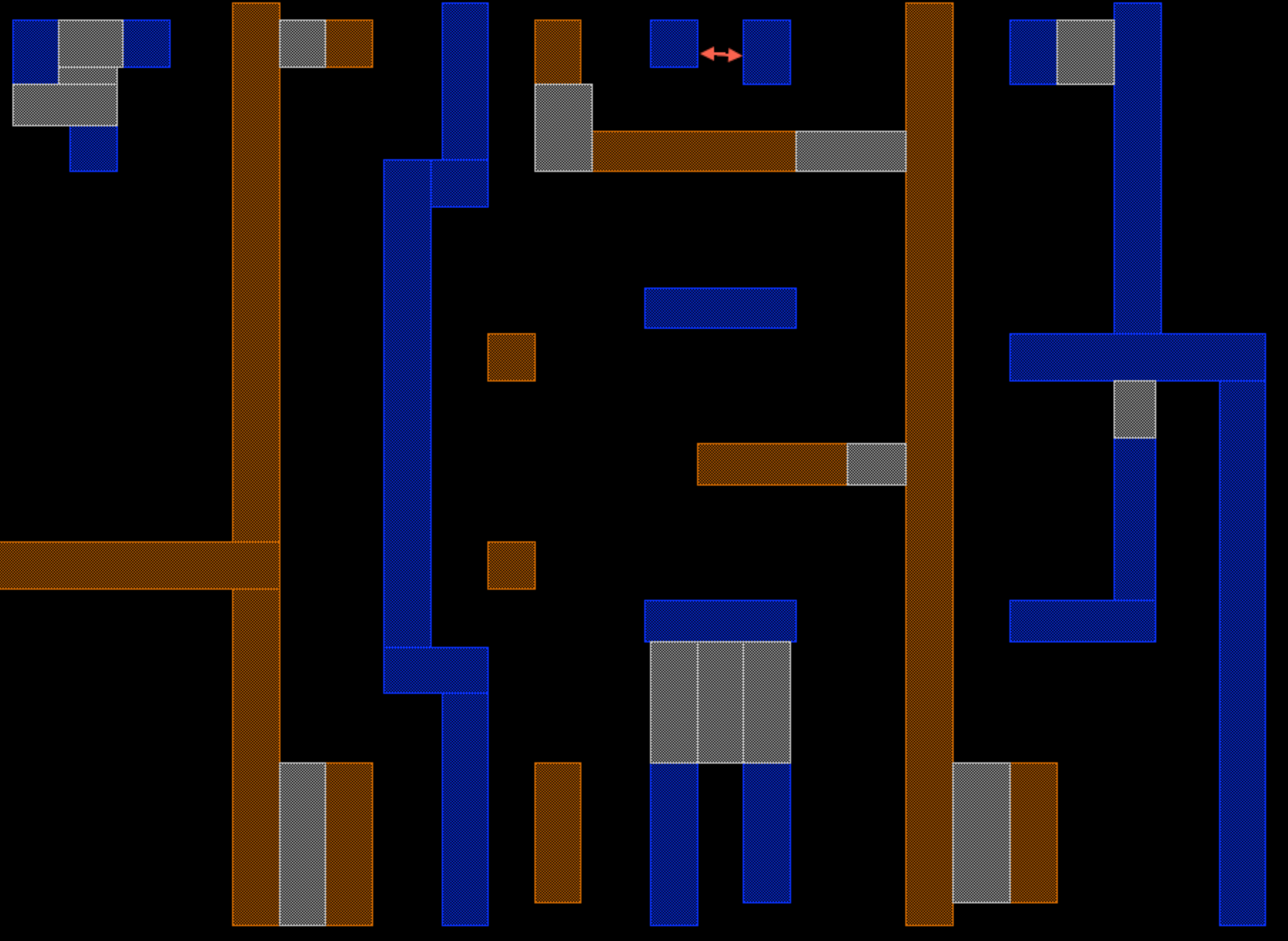}}
  \caption{Conflict examples in decomposed results.}
  \label{fig:result}
\end{figure}

Fig. \ref{fig:result} shows two conflict examples in decomposed layout, where conflict pairs are labeled with red arrows.
We can observe that both of the two conflicts come from via shapes.
One possible reason is that it is hard to find end-cut candidates around via, comparing with long wires.

\subsection{LELEEC v.s. LELELE}

\begin{figure}[htb]
  \centering
  \includegraphics[width=0.6\textwidth]{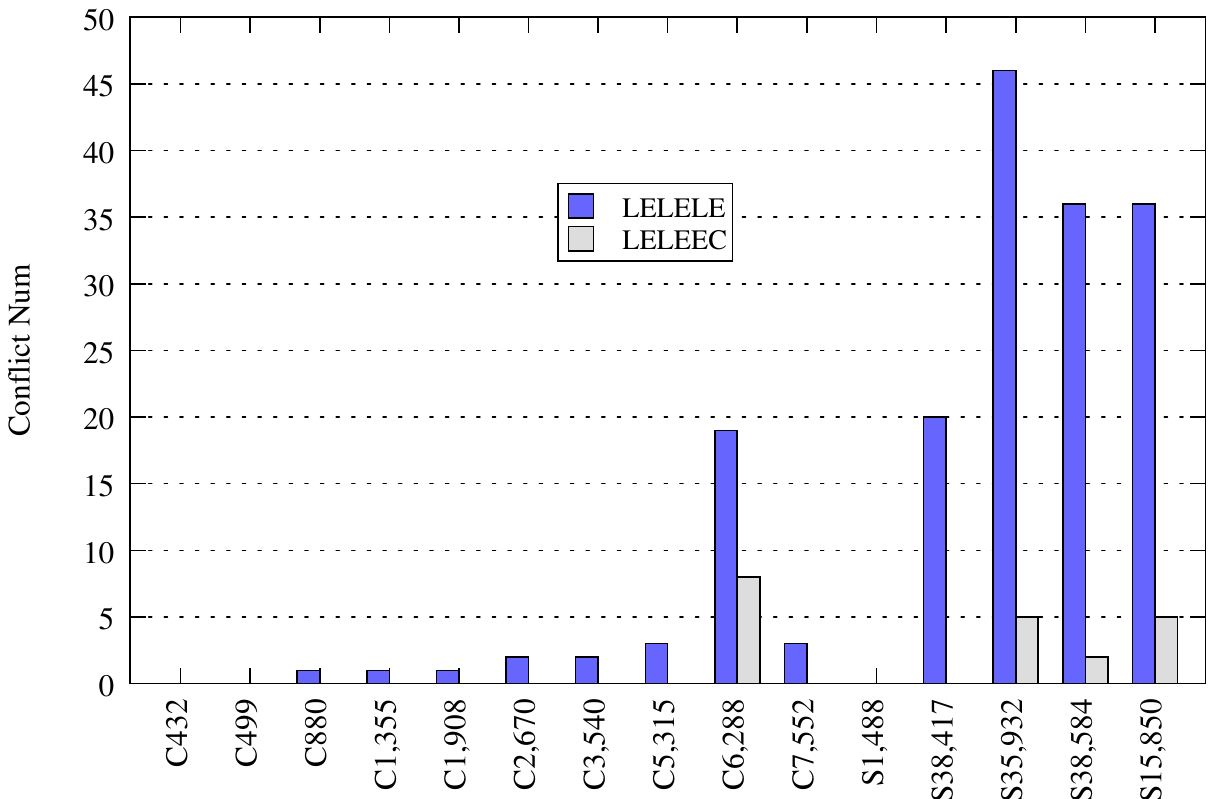}
  \caption{Conflict number comparison}
  \label{fig:conflict}
\end{figure}

\begin{figure}[htb]
  \centering
  \includegraphics[width=0.6\textwidth]{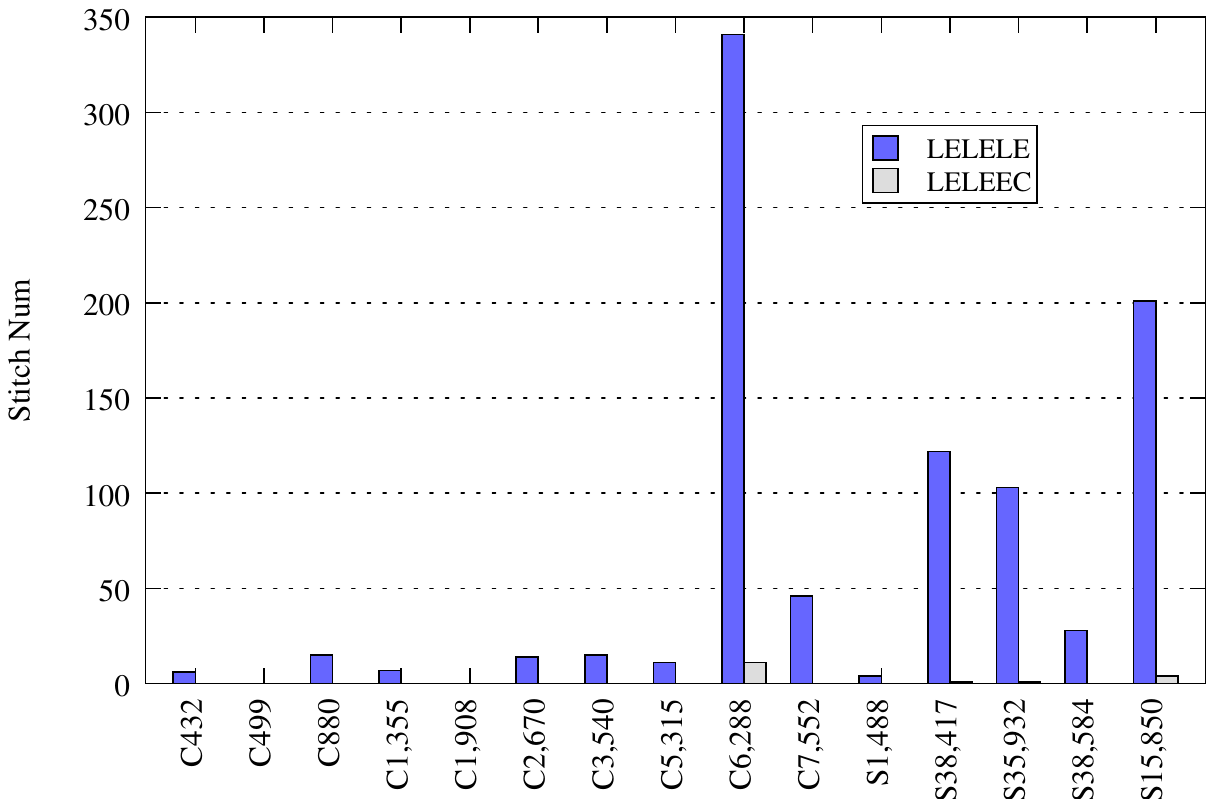}
  \caption{Stitch number comparison}
  \label{fig:stitch}
\end{figure}

In the next experiment, we compare two TPL type layout decompositions.
The state-of-art LELELE decomposer \cite{TPL_DAC2012_Fang} is applied for the comparison.
Fig. \ref{fig:conflict} and Fig. \ref{fig:stitch} compare the conflict number and the stitch number under two lithography processes, respectively.
We can observe that through applying end-cutting (LELEEC), both the conflict and stitch number are reduced dramatically.
The reasons are mainly twofold: (1) 4-clique conflict, which is a common type in standard layout, can be resolved in LELEEC; (2) Compared with wire shapes, most of the end-cuts are smaller, therefore the trim mask can contain more features. 

\section{Conclusion}
\label{sec:conclu}

In this paper, we propose the first study for the LELEEC layout decomposition.
The problem is translated into an optimization problem on the layout graph and end-cut graph.
Integer linear programming (ILP) is then applied to search the solutions.
The experimental results show the effectiveness of our algorithms.
In addition, our preliminary results show that compared with traditional LELELE type TPL,
LELEEC can reduce both the conflict number and stitch number dramatically.
As LELEEC may be adopted by industry for 14nm/11nm nodes, we believe more research will be needed to enable LELEEC-friendly design and mask synthesis.

\section*{Acknowledgment}

This work is supported in part by NSF, SRC, Oracle, and NSFC.

\bibliographystyle{IEEEtran}
\bibliography{./Ref/Bei,./Ref/Algorithm,./Ref/MPL,./Ref/Lith,./Ref/EBL,./Ref/EUV}

\end{document}